\documentclass[onecolumn,amsmath,amssymb,superscriptaddress,aps,prc]{revtex4-1}

\usepackage{graphicx}
\usepackage{epsf}
\usepackage{amsmath,amssymb}
\usepackage{bm}
\usepackage{color}
\usepackage{ulem}
\usepackage{stackengine}
\usepackage{braket}
\usepackage{multirow}

\newcommand{\figref}[1]{Fig.~\ref{fig:#1}}
\newcommand{\equref}[1]{Eq.~(\ref{eq:#1})}

\begin{document}

\title{
Efimov states in excited nuclear halos
}%
\author{Shimpei Endo}
\email{shimpei.endo@nucl.phys.tohoku.ac.jp}
\affiliation{Department of Physics, Tohoku University, Sendai 980-8578, Japan}
\affiliation{Department of Engineering Science, University of Electro-Communications, Chofu, Tokyo 182-8585, Japan}
\author{Junki Tanaka}
\email{junki@rcnp.osaka-u.ac.jp}
\affiliation{Research Center for Nuclear Physics, Osaka University, 10-1 Mihogaoka, Ibaraki, Osaka 567-0047, Japan}
\affiliation{RIKEN Nishina Center for Accelerator-Based Science, Wako, Saitama 351-0198, Japan}

\date{\today}

\begin{abstract}
Universality --- an essential concept in physics --- implies that different systems show the same phenomenon and can be described by a unified theory. A prime example of the universal quantum phenomena is the Efimov effect, which is the appearance of multiples of low-energy three-body bound states with progressively large sizes dictated by the discrete scale invariance. The Efimov effect, originally proposed in the nuclear physics context, has been observed in cold atoms and $^4\mathrm{He}$ molecules. The search for the Efimov effect in nuclear physics, however, has been a long-standing challenge owing to the difficulty in identifying ideal nuclides with a large $s$-wave scattering length; such nuclides can be unambiguously considered as Efimov states. Here, we propose a systematic method to identify nuclides that exhibit Efimov states in their excited states in the vicinity of the neutron separation threshold. These nuclei are characterised by their enormous low-energy neutron capture cross-sections, hence giant $s$-wave scattering length. Using our protocol, we identified $^{90}$Zr and $^{159}$Gd as novel candidate nuclides that show the Efimov states. They are well inside the valley of stability in the nuclear chart, and are suited for experimental realisation of the Efimov states in nuclear physics.

\end{abstract}

\maketitle

\begin{figure*}[!t]
	\centering
	\includegraphics[width=1.03\linewidth]{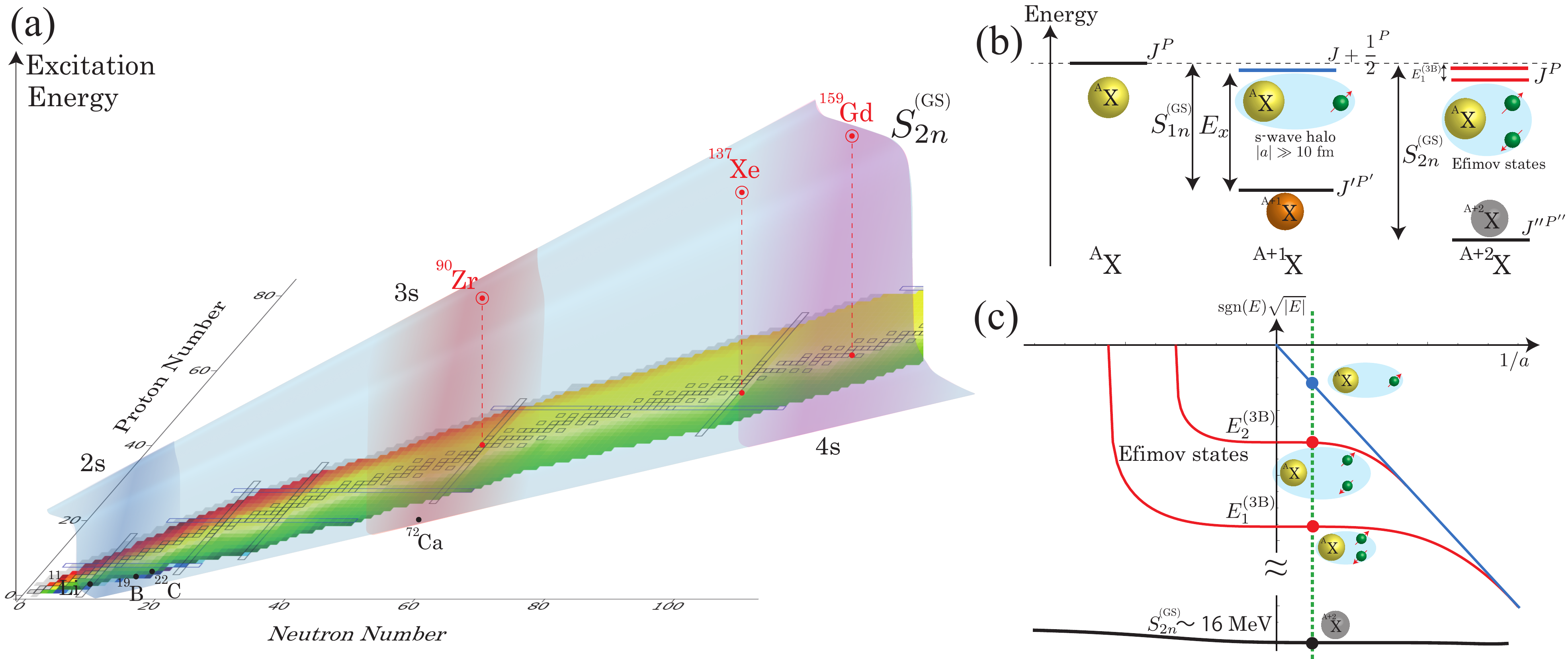} 
\caption{Schematic illustration of the Efimov states in the excited nuclei. (a) Excited states of the nuclei in the vicinity of two-neutron separation threshold $S_{2n}^{(\mathrm{GS})}$ (light blue 3D surface) are proposed to show the Efimov states. The near-threshold excited states of $^{90}$Zr and $^{159}$Gd (red circles) are specific candidates, whose ground states are well inside the valley of stability in the nuclear chart (National Nuclear Data Centre). These nuclei have outmost 3s (red) and  4s orbitals (purple), respectively. The proton-rich side is not shown because $S_{2n}^{(\mathrm{GS})}$ are complicated, and is out of scope of this paper. (b) The $s$-wave halo state of $\displaystyle _Z^{A}X$ and a neutron $n_{\uparrow}$, appearing as the excited state of $\displaystyle _Z^{A+1}X$ nuclei (blue level) in the vicinity of the neutron separation threshold (black dashed line), is the crucial ingredient for Efimov physics. It is in $J+1/2^{P}$, in conjunction with the $J^P$ state of $\displaystyle _Z^{A}X$, with very small two-body energy $S_{1n}^{(\mathrm{GS})}-E_x$, which can be estimated from one-neutron separation energy $S_{1n}^{(\mathrm{GS})}$ and the excitation energy $E_x$ as measured from the ground state. The Efimov states appear as excited states of $\displaystyle _Z^{A+2}X$ (red levels) in the vicinity of two-neutron separation threshold $S_{2n}^{(\mathrm{GS})}$ with $J^P$, the same spin-parity as that of $\displaystyle _Z^{A}X$ but distinct from the ground state of $\displaystyle _Z^{A+2}X$ (black bottom level). The rest mass of the neutrons is subtracted in these diagrams. (c) Schematic illustration of the energy spectrum as the $\displaystyle _Z^{A}X$-neutron $s$-wave scattering length (horizontal axis) is varied. Energy is measured from the neutron separation threshold (black dashed line in (b)). The weakly bound Efimov states (two red curves) and the tightly bound ground state (black curve) corresponds to the excited states (red levels) and the ground state (black level) in the right most of the level diagram of $\displaystyle _Z^{A+2}X$ in (b), whereas the dimer curve (blue curve) corresponds to the $s$-wave halo state (blue level) of $\displaystyle _Z^{A+1}X$ in (b). The physical value of the core-neutron $s$-wave scattering length (green vertical dotted line) depends on nuclides, which can be estimated from thermal neutron capture cross-section $\sigma_{\mathrm{TNC}}$ or $S_{1n}^{(\mathrm{GS})}-E_x$ data.}
\label{fig:NuChatMainFig}
\end{figure*}

At low energies, seemingly different quantum systems show the same behaviours independent of the microscopic details and can be described by the universal theory, as in the quantum phase transitions~\cite{sachdev1999quantum}. The Efimov states~\cite{efimov1970energy,naidon2017efimov,Braaten2006259,RevModPhys.89.035006}, weakly bound three-body bound states featuring the discrete scale invariance (\figref{NuChatMainFig}~(c)), are prime examples of the universal quantum phenomena. The Efimov states and their discrete scale invariance are observed in cold atoms~\cite{kraemer2006evidence,naidon2017efimov,RevModPhys.89.035006,PhysRevLett.112.190401,PhysRevLett.112.250404,PhysRevLett.113.240402}. Interestingly, the Efimov states observed in different atomic species and spin states are found to behave universally~\cite{PhysRevLett.107.120401,gross2011study,PhysRevLett.108.263001,pascaleno3BP1,pascaleno3BP2}. The Efimov states are also predicted to appear in a condensed matter system of quantum spins with the same discrete scale-invariant pattern~\cite{nishida2013efimov}. A key for the universality of the Efimov states is a large $s$-wave scattering length, which can be controllably realized in cold atoms with the Feshbach resonance~\cite{inouye1998observation,chin2010feshbach}.

In contrast, the Efimov states in nuclear physics have remained elusive. The halo states with small binding energy and large spatial size appearing in the neutron-rich nuclei near the neutron drip-line~\cite{tanihata20132013prpnp,PhysRevLett.73.2817,PhysRevC.56.R2378,AnnRev_HamPlatt,naidon2017efimov,FREDERICO2012939,canham2008universal,Hammer_2017,ACHARYA2013196} have been deemed as major candidates to show Efimov-like states; if a  neutron-rich nuclei $\displaystyle _Z^{A+1}X$ has a one-neutron halo structure with the neutron predominantly occupying an outmost $s$-orbital, the $\displaystyle _Z^{A+2}X$ nuclei may have a two-neutron halo structure of the Efimov characters. Indeed, the halo state of $\displaystyle _Z^{A+1}X$ may be considered as a two-body state of the $\displaystyle _Z^{A}X$ core and a neutron if their spatial extent is much larger than the radius of $\displaystyle _Z^{A}X$. If the neutron is in $s$ orbital, the core $\displaystyle _Z^{A}X$ and the neutron $n_\uparrow$ (assumed here to be in the spin-up state for later purposes) can be considered as a two-body system with a large $s$-wave scattering length $a$. The halo $\displaystyle _Z^{A+1}X$ nucleus can then be universally described with $a$:
\begin{equation}
\label{eq:dimEneuniva}E_{2\mathrm{B}}=-\hbar^2/2\mu a^2, \ \ \ \ \  \sqrt{\langle r^2 \rangle}=a/\sqrt{2},
\end{equation}
where $\mu $ is the reduced mass between $\displaystyle _Z^{A}X$ and the neutron. The two-body energy $E_{2\mathrm{B}}$ corresponds to the energy of the $\displaystyle _Z^{A+1}X$ nucleus measured from the $\displaystyle _Z^{A}X + n_\uparrow$ separation threshold, and the mean-square radius $ \sqrt{\langle r^2 \rangle}$ is the distance between $\displaystyle _Z^{A}X$ and $n_\uparrow$. Because of the large $s$-wave scattering length between $\displaystyle _Z^{A}X$ and $n_\uparrow$, and that between the neutrons in the spin $\uparrow$ and $\downarrow$ states (i.e.$^1S_0$ state), we can consider a system of $\displaystyle _Z^{A}X$ nucleus with the two neutrons $n_\uparrow$ and $n_\downarrow$ as a three-body system interacting with large $s$-wave scattering lengths. Such a mass-imbalanced three-body system show the Efimov states when the $s$-wave scattering length is large. As $|a|$ gets larger, they tend to follow the discrete scale invariance, with a scale factor $e^{\pi/|s|}\approx 16-20$ universally determined by the mass ratios between the particles~\cite{Braaten2006259,naidon2017efimov,RevModPhys.89.035006}. Although this scenario seems plausible, finding a neutron-rich nucleus that unambiguously shows the universal Efimov features is challenging; the neutron-rich halo nuclei are often not in the low-energy regime to be described by the universal Efimov theory. The finite-range effects of the nuclear forces are relevant. and the higher orbital angular-momentum channels play crucial roles in some nuclei, such as $^{11}$Li~\cite{tanihata1985measurements,PhysRevC.50.R550}. Furthermore, the neutron-rich nuclei are not easy to synthesise especially for extreme neutron-rich nuclei, such as $^{62}$Ca~\cite{PhysRevLett.111.132501,PhysRevC.105.024310}, $^{72}$Ca~\cite{PhysRevLett.120.052502}. With these challenges, there are only a handful of candidates for the nuclear Efimov states. For most of them, however, the $s$-wave scattering length is at most 10 fm order. The best candidate to realize universal Efimov physics with the current experimental technique is $^{19}$B~\cite{PhysRevLett.121.262502,PhysRevC.100.011603}, whose scattering length is $\gtrsim 100$ fm.

Here, we propose an alternative route to search for nuclear systems with extremely large $s$-wave scattering lengths, and three-body bound states that can distinctly be identified as Efimov states. In contrast to neutron-rich nuclei, we focus on nuclei well inside the valley of stability in the nuclear chart  (\figref{NuChatMainFig}~(a)). While their ground states are stable with a large binding energy per nucleon  $\sim 8$ MeV, we focus on their excited state in the vicinity of the neutron separation threshold. Suppose the excitation energy $E_x$ of the nucleus $\displaystyle _Z^{A+1}X$ is very close to the one-neutron separation energy $S_{1n}^{(\mathrm{GS})}$. We may describe the excited state of $\displaystyle _Z^{A+1}X$ as a weakly bound two-body state of $\displaystyle _Z^{A}X$ nucleus and a neutron $ n_{\uparrow}$ with its relative two-body energy $S_{1n}^{(\mathrm{GS})}-E_x$ (centre of \figref{NuChatMainFig}~(b)). If this halo state mainly occupies an $s$-wave orbit, $\displaystyle _Z^{A+1}X$ can be considered as a two-body system with a large $s$-wave scattering length. If we then add another  neutron with an opposite spin, we have a three-body system $\displaystyle _Z^{A}X$-$n_{\uparrow}$-$n_{\downarrow}$ interacting with large $s$-wave scattering lengths, which is expected to show the Efimov states, as an `Efimov favoured system'~\cite{Braaten2006259,naidon2017efimov}. With smaller $|S_{1n}^{(\mathrm{GS})}-E_x|$, the $s$-wave scattering lengths are larger (see Eq.~(\ref{eq:dimEneuniva}))). As shown later, there are nuclear excited states with $|S_{1n}^{(\mathrm{GS})}-E_x|\sim10$ meV, corresponding to $|a| \sim 10^4$ fm. Because the number of Efimov states increases as $|a|$ increases and show clearer universal features~\cite{efimov1970energy,Braaten2006259,naidon2017efimov}, we expect more Efimov states with better universal features for such nuclear systems; for example, they behave as Borromean states for $1/a<0$ and breaks up into a halo dimer state of $\displaystyle _Z^{A}X$ and a neutron for $1/a>0$ as the inverse scattering length between $\displaystyle _Z^{A}X$ and a neutron $1/a$ is continuously varied (\figref{NuChatMainFig}~(c)). Because the Efimov states are weakly bound, the Efimov states in $\displaystyle _Z^{A+2}X$ nucleus should appear as the excited states in the vicinity of the two-neutron separation threshold (the right most of \figref{NuChatMainFig}~(b)). The spin-parity $J^P$ of the Efimov states is the same as that of core $\displaystyle _Z^{A}X$ because the anti-parallel neutrons in the $s$-wave orbit (parity $0^+$) couples to the core.  Consequently, as summarized in \figref{NuChatMainFig}~(b), the Efimov states in $\displaystyle _Z^{A+2}X$ appear as sub-threshold excited states with $J^P$, along with a sub-threshold excited state of $\displaystyle _Z^{A+1}X$ with $J+1/2 ^{P}$, adjacent to the $\displaystyle _Z^{A}X$ ground-state nucleus with $J^{P}$. The Efimov states are distinct from the ground-state of $\displaystyle _Z^{A+2}X$, which has a large binding energy per nucleon $\sim 8$ MeV and can have angular momentum and parity $J''^{P''}$ distinct from the Efimov states $J^P$. Similarly, the ground state of $\displaystyle _Z^{A+1}X$ in $J'^{P'}$ is distinct from the excited halo state in $J+1/2 ^{P}$. The feature of the Efimov states having the same $J^{P}$ as the core because of the anti-parallel neutrons sounds akin to the two-neutron pairing. However, the neutron pairing with a conventional BCS-like mechanism is a many-body phenomenon requiring a finite background neutron density, occurring inside or at the surface of the core nucleus. Contrarily, the Efimov states are few-body phenomena that occur far away from the core. Further, the neutron pairing can occur even when the neutrons occupy a non-$s$-wave orbit such that the resulting states can take various angular-momentum values allowed by the addition rule, while the Efimov states are restricted to the same $J^{P}$ as the core.

\section{\label{sec:TNC_2}Diagnosis of large $S$-wave scattering length with neutron capture cross-section}
One naive way to find good candidates that would show the Efimov states in their sub-threshold excited states is to look for the nuclear data of $S_{1n}^{(\mathrm{GS})}$ and $E_x$ and search for nuclei with small $|S_{1n}^{(\mathrm{GS})} - E_x| \ll $ 1~MeV. Although this approach provides a lot of seemingly good candidates, most of them do not show the Efimov states; this is because the sub-threshold states often do not exhibit halo $s$-wave characteristics; it can be a conventional nuclear excited state coincidentally appearing around the neutron separation threshold; it can be a low-energy collective excitation, such as giant resonances; even if it is a halo state of $\displaystyle _Z^{A}X$ and a neutron, it can be in the non-$s$-wave orbitals. Therefore, the $|S_{1n}^{(\mathrm{GS})} - E_x| $ values would give us too many `spurious' candidates.

\begin{figure}[!t]
	\centering
	\includegraphics[width=0.4\linewidth]{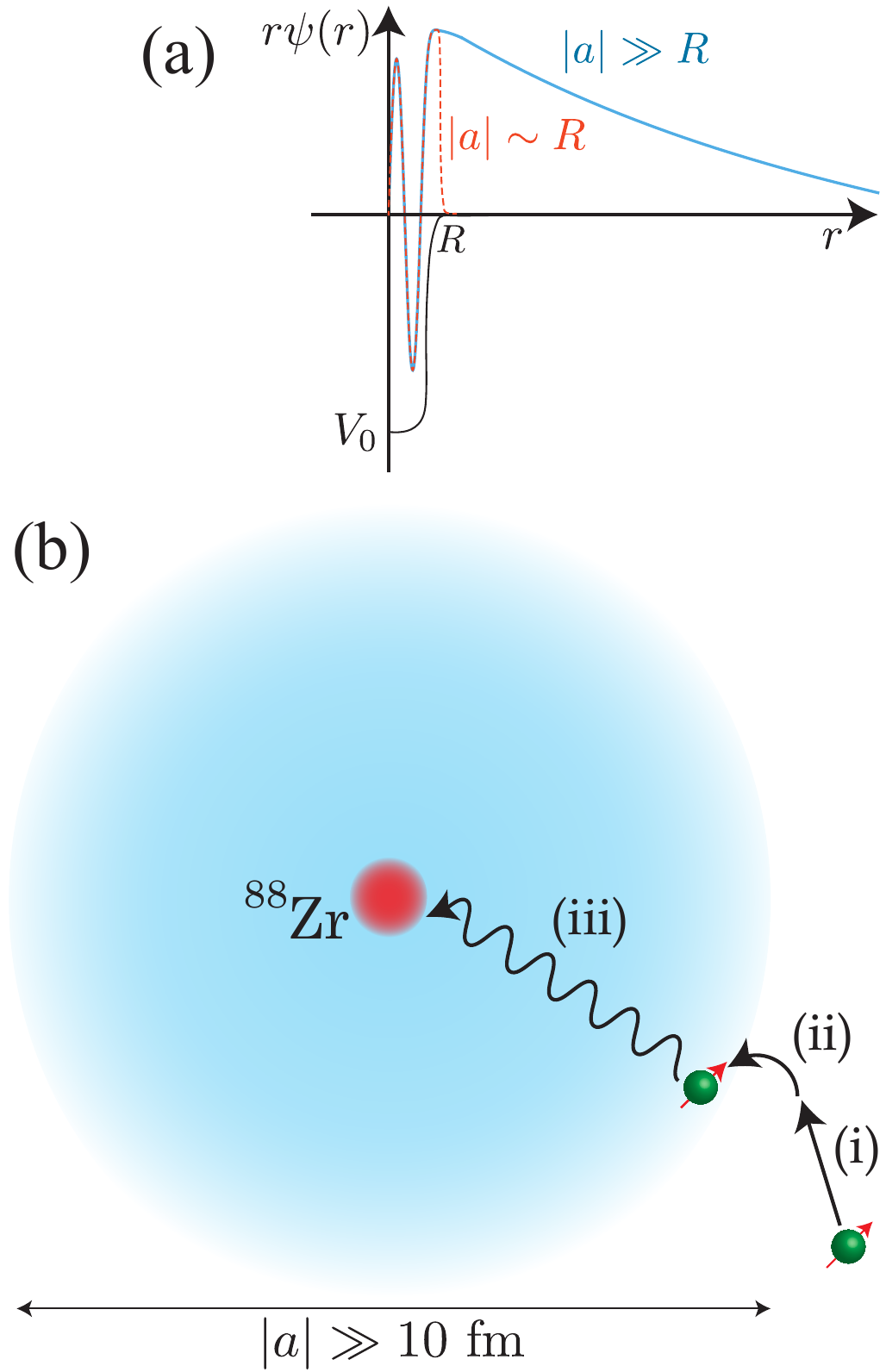} 
\caption{(a) When the $s$-wave scattering length $a$ is much larger than the typical range of the interaction $R$, $|a| \gg R$, there is a halo state (virtual state for $a>0$ and bound state for $a<0$), whose wavefunction between the nucleus core and a neutron $\psi$ extends up to a distance much larger than $R$ (solid curve). This is in sharp contrast with a typical bound state with $|a|\sim R$ (orange dotted curve), whose spatial size is as small as $R$. (b) Schematic illustration of the thermal neutron capture process via the halo state. While $\sigma_{\mathrm{TNC}}\sim \pi R^2\sim 1$-$10\ \mathrm{fm}^2$ for an ordinary process in which a neutron is directly captured by the core nucleus, in a resonant capture process, the low-energy neutron capture cross-section is enhanced by a halo state around the core as follows: (i) An incident neutron, typically at much smaller energy than the nuclear energy scales, approaches to distance $r\sim |a|$. (ii) For typical nuclei, the neutron cannot get captured by the nucleus (red region) as the neutron is far away from the nucleus with its size $R\ll r$. Because of the $s$-wave halo state with a very large spatial size (blue region), the neutron can be virtually captured into the halo state, even when it is much further away from the core. (iii) The captured neutron may, after a while, decay into an energetically more stable state by emitting energy, e.g. multiple $\gamma$-rays. Thus, the halo state acts as a  \textit{doorway} state, through which the neutron capture rate is significantly enhanced.}
\label{fig:TNCSchematic}
\end{figure}

Herein, we proposes that the thermal neutron capture cross-section provides much more direct access to the nuclei with the excited Efimov states. The thermal neutron capture is a reaction where the target nucleus captures an incident neutron, whose rate is represented by a cross-section $\sigma_{\mathrm{TNC}}$. If the incident energy is small, as is the case for experiments performed at meV to eV~\cite{shusterman2019surprisingly,bjerrum1960low}, the $s$-wave capture should be dominant. The incident neutron would silently come in close contact with the static core, rather than abruptly exciting its collective motions. This expectation should be particularly true when there is a weakly bound $s$-wave halo state: a neutron can be captured  through such a \textit{doorway state} without affecting much the core nucleus (\figref{TNCSchematic}(b)). The capture cross-section gets larger as the radius of the \textit{doorway state} increases. We can therefore expect that nuclei with a large $\sigma_{\mathrm{TNC}}$ are likely to possess a sub-threshold $s$-wave state with small binding energy and large spatial size, hence large $s$-wave scattering length (see \equref{dimEneuniva}). Using the conventional formula of the low-energy elastic scattering $\sigma_{\mathrm{el}}= \pi a^2$ and assuming that it would be the same order of magnitude as the neutron capture cross-section as is the case for the `black-body' scattering~\cite{Satchler_G_R_1990-02-05}, we obtain
\begin{equation}
\label{eq:TNCaEstimate} |a| \approx \sqrt{\frac{ \sigma_{\mathrm{TNC}}}{\pi} }
\end{equation}
as a crude but useful estimate of the $s$-wave scattering length.

As notable examples, extremely large neutron capture cross-sections have been reported:  $\sigma_{\mathrm{TNC}} =8.27(64)\times 10^5 $ b for $^{88}$Zr~\cite{shusterman2019surprisingly}, and $\sigma_{\mathrm{TNC}} =2.54(2)\times 10^5 $ b for $^{157}$Gd~\cite{bjerrum1960low,shusterman2019surprisingly}, respectively. From \equref{TNCaEstimate}, the $s$-wave scattering lengths for $^{88}$Zr-$n$ and $^{157}$Gd-$n$ are estimated to be $|a|=5.1(2) \times 10^3$~fm and $|a|=2.84(1) \times 10^3$ fm respectively.  These values are about two orders of magnitude larger than those in the neutron-rich nuclei. Therefore, we expect that $^{88}$Zr and $^{157}$Gd plus two neutrons, namely the excited states in the $^{90}$Zr and $^{159}$Gd, are three-body systems with very large $s$-wave scattering lengths, which are conductive to observing clear signatures of Efimov states. From the $S_{1n}^{(\mathrm{GS})}$ and $E_x$ data, we can further strengthen the evidence for the Efimov states. Indeed, $^{89}$Zr possesses a sub-threshold state with an extremely small energy $|S_{1n}^{(\mathrm{GS})}-E_x| \le  25 $ meV~\cite{PhysRevC.103.024614,ZrBEneRef1}, too small to identify whether it is a bound ($S_{1n}^{(\mathrm{GS})}-E_x >0$) or virtual state ($S_{1n}^{(\mathrm{GS})}-E_x <0$). In either case, we can estimate the $s$-wave scattering length of  $^{88}$Zr-$n$ as (\equref{dimEneuniva} valid even for a virtual state) $|a| \ge 2.9\times 10^4$~fm. Although this value is a few times different from that estimated from $\sigma_{\mathrm{TNC}}$ because \equref{TNCaEstimate} is a crude estimate, the two different physical quantities both suggest enormous $s$-wave scattering lengths. The same is true for Gd, i.e., $^{158}$Gd has a virtual state with $S_{1n}^{(\mathrm{GS})}-E_x= 31.4$ meV~\cite{bjerrum1960low,10.1093/ptep/ptz002}. This suggests a large negative $s$-wave scattering length $ a = -2.6\times 10^4$~fm. $^{88}$Zr and $^{157}$Gd should therefore be good candidates for the Efimov states in the sub-threshold excited states. These are not the only candidates; with further neutron capture measurements, other nuclei with large $a$ may be discovered.


Some remarks are in order. First, the $s$-wave scattering length of the core and a neutron should be simultaneously large for $n_{\uparrow}$ and $n_{\downarrow}$. If this condition is not satisfied, then one of the $s$-wave scattering length may take a naturally small value  $|a|\sim  1$ fm. The excited states of $\displaystyle _Z^{A+2}X$ would then be a three-body system with only two resonant interactions. It corresponds to an `Efimov unfavoured system'~\cite{Braaten2006259,naidon2017efimov}, where the binding energies are too small and the discrete scale factor is too large to observe in experiments. It is, therefore, crucial to have the same or approximately the same $s$-wave scattering lengths for $\displaystyle _Z^{A}X$-$ n_{\uparrow}$ and $\displaystyle _Z^{A}X$-$ n_{\downarrow}$. This is true for Zr: $^{88}$Zr is in $J^P = 0^+$, dictating that the $s$-wave scattering lengths of $^{88}$Zr-$n_\uparrow$ and $^{88}$Zr-$n_\downarrow$ are the same and simultaneously large. By contrast, $^{157}$Gd is in $J^P=3/2^-$, so that its scattering with $n_{\uparrow}$ and $n_{\downarrow}$ can be different. Because $\sigma_{\mathrm{TNC}}$ is enormous, at least one of the scattering length should be large. However, with current nuclear data, we cannot conclude whether or not $^{157}$Gd-$n_{\uparrow}$ ($J=2^-$) and $^{157}$Gd-$n_{\downarrow}$ ($J=1^-$) both have large $s$-wave scattering lengths. Indeed, the spin-parity of the sub-threshold state of $^{158}$Gd is reported as $J^P=2^-,1^-$~\cite{greenwood1978collective}, and there is no clear knowledge on the relative strengths of the interactions. Therefore, as a working hypothesis, we suppose in the following that the $s$-wave scattering lengths of $^{157}$Gd-$n_{\uparrow}$ and $^{157}$Gd-$n_{\downarrow}$ are the same.

Second, the above argument of the thermal neutron capture is not rigorous and there are nuclei in which the neutron capture is enhanced by other factors (e.g. fission of U nucleus). Nevertheless, our approach is suitable for efficiently pinning down possible candidate nuclei. The other approach based on $S_{1n}^{(\mathrm{GS})}$ and $E_x$ is often contaminated by spurious states of complicated collective nature or higher orbitals. Furthermore, $S_{1n}^{(\mathrm{GS})}$ and $E_x$ near the neutron separation threshold often gets blurred by a high level density or broad experimental widths. The neutron capture cross-section would give us a simple but useful guideline for which nuclei may have a possible sub-threshold state of a halo nature. This is exemplified by $^{42-44}$Ca: $^{43}$Ca nucleus has an excited state $E_x=7.93270(3)$ very close to $S_{1n}^{(\mathrm{GS})}=7.93289(17)$ MeV~\cite{gruppelaar1969investigation}. It is then  tempting to expect an $s$-wave halo of $^{42}$Ca and a neutron with an extremely small binding energy $S_{1n}^{(\mathrm{GS})}-E_x = 0.19(17)$ keV and large $s$-wave scattering length  $a=340\left(
\begin{array}{c}
+760 \\
-100
\end{array}
\right)$ fm. However, $\sigma_{\mathrm{TNC}}$ of  $^{42}$Ca is small. This is because the sub-threshold state of  $^{43}$Ca is not well described by
 the neutron plus $^{42}$Ca core picture; it may be a complicated superposition of various states including collective excitations of $^{43}$Ca. Our approach can efficiently exclude such nuclei with dominant non-halo contributions.

\section{Efimov state in the excited nuclei}

We perform three-body calculations to demonstrate Efimov physics in the excited states of $^{90}$Zr and $^{159}$Gd (see Methods for details). The nuclei  $^{88}$Zr and $^{157}$Gd are modelled as single particles interacting with the two neutrons by the Woods--Saxon potentials. The interaction strength is varied around the third and fourth $s$-wave resonances, which correspond to the dominant 3s and 4s orbitals around the thresholds in $^{90}$Zr and $^{159}$Gd, respectively. The interaction between the neutrons is modelled by the AV4 potential with $a_{nn} = -19.5$~fm~\cite{PhysRevLett.89.182501}. Our three-body model is particularly suited for low-energy sub-threshold states, where their size is large and the core excitations is irrelevant, whie it cannot describe their ground states owing to their large binding energies ($\sim 16$~MeV). To compare it with a hypothetical neutron-neutron interaction at the unitary limit $a_{nn} = \pm \infty$, an ideal situation for the Efimov states, we also consider a scaled neutron-neutron potential $\lambda V_{nn}(r)$,  where $\lambda$ is fine-tuned $\lambda=1.07...$ to be $a_{nn} = \pm \infty$.



\begin{figure*}[!t]
	\centering
	\includegraphics[width=\linewidth]{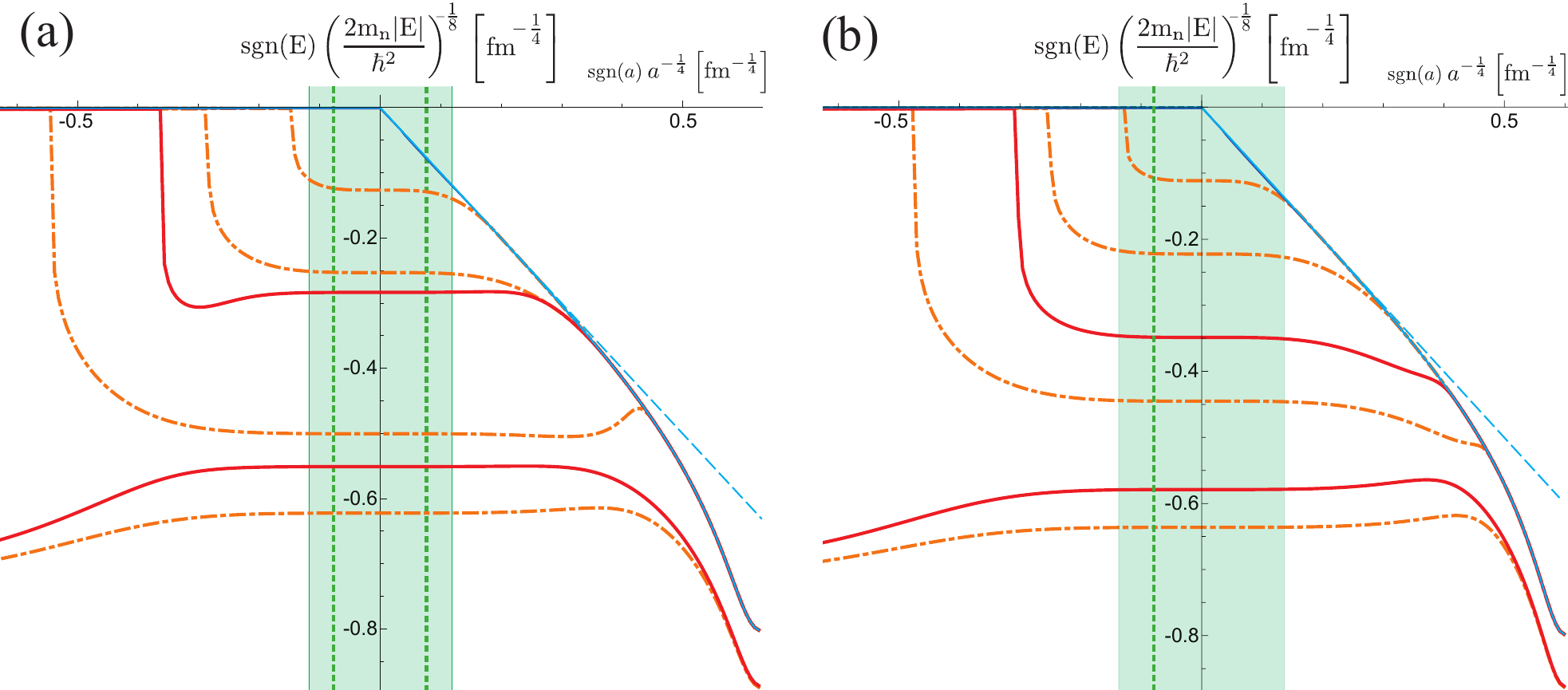} 
\caption{Energy spectra of the three-body systems for variable core-neutron $s$-wave scattering length $a$ for (a) $^{88}$Zr+2n, and (b)$^{157}$Gd+2n. The zero energy is set at the three-body threshold energy; e.g. $^{88}$Zr+n+n threshold energy (with the rest mass of the neutron subtracted), i.e. the energy of the $^{88}$Zr nucleus. The thick solid curves (red) correspond to the physical neutron-neutron potential, AV4 potential, with $a_{nn} = -19.5$~fm, and the thick dashed-dotted curves (orange) correspond to a fictitious neutron-neutron potential, namely scaled AV4 potential tuned to the unitary limit $1/a_{nn} = 0$ (we only show up to the fourth trimer states). The thin curves (blue) are two-body energy of the core nucleus + one neutron, and the thin dashed lines (blue) are the universal dimer energy in \equref{dimEneuniva}. The green regions and the green vertical dotted lines are the physical $s$-wave scattering length of the nucleus and a neutron $a$ estimated from the thermal neutron capture cross-section data (\equref{TNCaEstimate}) and the neutron separation energy data (\equref{dimEneuniva}), respectively: (a) As $|S_{1n}^{(\mathrm{GS})}-E_x| \le  25 $ meV~\cite{PhysRevC.103.024614,ZrBEneRef1} and the sign is unknown for Zr, we denote two vertical dotted lines, each corresponding to $a = \pm 2.9\times 10^4$~fm. (b) For Gd, $S_{1n}^{(\mathrm{GS})}-E_x= 31.4$ meV~\cite{bjerrum1960low,10.1093/ptep/ptz002}, which suggests a large negative $s$-wave scattering length $ a = -2.6\times 10^4$~fm denoted as a vertical dotted line.}
\label{fig:AllSpeciesPhysicalRes}
\end{figure*}


In \figref{AllSpeciesPhysicalRes}, we show the three-body energies $E_n^{(\mathrm{3B})}$ for (a)$^{88}$Zr+2n, and (b)$^{157}$Gd+2n, respectively. When the neutron-neutron interaction is strictly at the unitary limit $a_{nn} = \pm \infty$ (orange dashed-dotted curves), an infinite tower of the Efimov states appear around the unitary limit of the core-neutron scattering length $1/a =0$. In particular, the excited states excellently show the universal Efimov features; start to appear as Borromean states for $1/a<0$ and dissociate into a particle and a dimer for $1/a>0$ as $1/a$ is varied. They also show the discrete scale-invariant pattern dictated by the Efimov theory: the ratios of the binding energies at the unitary limit $1/a=0$ are $\sqrt{E_{n+1}^{(\mathrm{3B})}/E_{n}^{(\mathrm{3B})}}=15.26(5), 15.95(5) $ for Zr and $\sqrt{E_{n+1}^{(\mathrm{3B})}/E_{n}^{(\mathrm{3B})}}=16.05(5), 15.86(5)  $ for Gd, which are in excellent agreement with the universal values of the Efimov theory $e^{\pi/|s|}=15.95...$ and 15.86..., respectively. Contrarily, the ground state in \figref{AllSpeciesPhysicalRes} exhibit marginal Efimov features; it neither dissociates into three particles for $1/a<0$ nor into a particle plus a dimer for $1/a>0$, and the binding energies ratios at the unitary limit are $\sqrt{E_{2}^{(\mathrm{3B})}/E_{1}^{(\mathrm{3B})}}= 2.4(1) $ for Zr and $\sqrt{E_{2}^{(\mathrm{3B})}/E_{1}^{(\mathrm{3B})}}=4.2(2)$ for Gd. The ground state with a large binding energy is affected by the finite-range effects and is less likely to show the Efimov behaviour dictated by the universal theory. Thus, the ground state with energy $E_1^{(\mathrm{3B})}=0.466(2)$~MeV (Zr, $1/a=0$) $E_1^{(\mathrm{3B})}=0.556(2)$~MeV (Gd, $1/a=0$)  is much more affected by the finite-range effects than the first excited state, which has much smaller binding energies $E_2^{(\mathrm{3B})}=81.6(4)$~keV (Zr) and $E_2^{(\mathrm{3B})}=31.9(1)$~keV (Gd). We note that these ground states in  \figref{AllSpeciesPhysicalRes} should not be confused with the true ground state of $^{90}$Zr and $^{159}$Gd; the binding energy of the ground states in  \figref{AllSpeciesPhysicalRes} $E_1^{(\mathrm{3B})}$ is considerably smaller than the true ground state binding energy $S_{2n}^{(\mathrm{GS})}\approx 16$~MeV, which cannot be described by our low-energy three-body model. The ground states in \figref{AllSpeciesPhysicalRes} should be considered as sub-threshold excited states of $^{90}$Zr and $^{159}$Gd.


In \figref{AllSpeciesPhysicalRes}, we also show the results with a realistic neutron-neutron interaction with $a_{nn} = -19.5$~fm (red solid curves). In contrast to the unitary one, we only find two three-body bound states because the realistic neutron-neutron interaction is slightly shallower than the unitary one. The excited trimer is extremely weakly bound $E_2^{(\mathrm{3B})}=0.868(4)$~keV (Zr $1/a=0$) and $E_2^{(\mathrm{3B})}=4.52$~keV (Gd $1/a=0$), with their sizes estimated as $\sqrt{\langle r^2 \rangle} \sim \sqrt{\hbar^2/2m_n|E_2^{(\mathrm{3B})}|}= 155$~fm for Zr and $68$~fm for Gd, where $m_n$ is the rest mass of the neutron. These states originate from the Efimov states in the case of $a_{nn} = \pm \infty$. In particular, the weaker bound states excellently show the Efimov features: they start to appear as Borromean states for $1/a<0$, follow the behaviours of the Efimov states for $1/a_{nn}=0$ for $1/a \approx 0$, and dissociate into a particle and a dimer (blue solid curve) for $1/a>0$ as $1/a$ is varied. Thus, the weaker bound state of Zr is akin to the second excited Efimov state of $a_{nn} = \pm \infty$, whereas that of Gd behaves similarly to the first excited Efimov state of $a_{nn} = \pm \infty$, pushed up because the neutron-neutron scattering length is not infinite. The tighter bound state shows relatively less clear Efimovian characters due to finite-range effects, but it still may be considered as the Efimov states for most regions; it behaves as a Borromean state for $1/a<0$ and the energy spectrum behaves in an analogous manner as $1/a$ is varied from the unitary limit toward $1/a>0$. The exceptions are $1/a\ll0$ and  $1/a\gg0$ where they neither dissociate into three particles nor a particle plus a dimer due to finite-range effects. Except for these regions, the weaker and tighter bound sub-threshold states of Zr and Gd seems highly likely to be Efimov-like states. The ratios between the tighter and weaker bound states are $\sqrt{E_{2}^{(\mathrm{3B})}/E_{1}^{(\mathrm{3B})}}= 14.2 $ for Zr and $\sqrt{E_{2}^{(\mathrm{3B})}/E_{1}^{(\mathrm{3B})}}=7.6(3)  $ for Gd at $1/a=0$; these values are slightly better than those at $1/a_{nn}=0$  (orange dashed-dotted curves) because of the smaller binding energies, which result in weaker finite-range effects.




By using the $\sigma_{\mathrm{TNC}}$ data, we can more specifically determine the binding energy of the Efimov trimers. From the $\sigma_{\mathrm{TNC}}$ data, we can estimate the absolute values of $a$ from \equref{TNCaEstimate}, which are represented as left-most and right-most boundaries of the green regions in \figref{AllSpeciesPhysicalRes}. Although these are not the accurate values of the $s$-wave scattering length, they are still useful for evaluating the order of magnitude of $a$ in the Efimov plot (shown by coloured regions in \figref{AllSpeciesPhysicalRes}) and estimating the binding energies of the sub-threshold trimer states as $E_1^{(\mathrm{3B})}= -176(1)$~keV  and $E_2^{(\mathrm{3B})}=-0.87(1)$~keV for  $^{90}$Zr and $E_1^{(\mathrm{3B})}=-261(2)$~keV  and $E_2^{(\mathrm{3B})} =-4.51(24)$~keV for $^{159}$Gd. Better estimates of the value and the sign of $a$ can alternatively be obtained if accurate data of $S_{1n}^{(\mathrm{GS})}$ and $E_x$ are available; the $s$-wave scattering length estimated from \equref{dimEneuniva}, dimer binding energy (blue curve), and $S_{1n}^{(\mathrm{GS})}-E_x$~\cite{PhysRevC.103.024614,ZrBEneRef1,bjerrum1960low,10.1093/ptep/ptz002} is denoted as vertical dotted lines (green), giving us the estimate $E_1^{(\mathrm{3B})}= -176(1)$~keV  and $E_2^{(\mathrm{3B})} =-0.868(2)$~keV for  $^{90}$Zr and $E_1^{(\mathrm{3B})}=-261.7(1)$~keV  and $E_2^{(\mathrm{3B})} =-4.49(2)$~keV for $^{159}$Gd. These values are in excellent agreement with those with $\sigma_{\mathrm{TNC}}$, vindicating our approach to search for the large $s$-wave scattering length nuclei with $\sigma_{\mathrm{TNC}}$.

To examine the dependence of these results on inter-particle interactions, we also perform our three-body calculations for different choices of neutron-neutron interactions (see Appendix for details). Most of the qualitative features are found to remain the same as \figref{AllSpeciesPhysicalRes}. One notable difference is that the presence of the weaker bound state is sensitive to the value of $a_{nn}$. We found that only the tighter bound trimer of $|E_1^{(\mathrm{3B})}|\sim 100$~keV appear for the neutron -neutron interaction with slightly smaller $|a_{nn}|$. While accurate inputs of $a_{nn}$ and $a$ are necessary to conclude the presence/absence of the second trimer, we universally find that at least one trimer with very small binding energy $E_1^{(\mathrm{3B})} \sim 100$~keV and large spatial size $\sqrt{\langle r^2 \rangle} \sim 10-20$~fm exist irrespective of specific choice of inter-nucleon potentials. Therefore, we conclude that at least one, and possibly two, Efimov trimers appear as sub-threshold excited states of $^{90}$Zr and $^{159}$Gd. As the neutron-neutron interaction is common for all nuclei, we may also conjecture that there can also be one or at best two Efimov states for the other candidates in the ridge stripes in \figref{NuChatMainFig}(a), owing to a finite value of $a_{nn}$. However, this hypothesis must be tested in future studies,

\section{`Ridge Stripes' for the Efimov states above the nuclear chart}
$^{90}$Zr and $^{159}$Gd are not the only nuclei, but should be a few examples of broader series of halo states. Let us give a global viewpoint on where in the nuclear chart our sub-threshold excited Efimov states should appear. Consider starting from light balanced nuclei and increase the number of nucleons along the valley of stability, i.e. diagonal direction in the nuclear chart. As the nucleon number increases, the nucleons in the ground-state nuclei progressively occupy higher $n l$ states as dictated by the shell model, which means an increase of the Fermi energy. The energy of the excited $ns$ states as measured from the Fermi energy, i.e. $E_x$ of $ns$ states, would therefore decrease. Conversely, the one-neutron separation energy for the stable nuclei inside the valley of stability is almost constant $S_{1n}^{(\mathrm{GS})}\sim 8$~MeV. Therefore, $E_x-S_{1n}^{(\mathrm{GS})}$ tends to decrease as the number of nucleons increases. We then arrive at the conjecture that there are optimal numbers of nucleons for each 2s, 3s, 4s states which realise small $E_x-S_{1n}^{(\mathrm{GS})}$. This suggests that there are regions in the nuclear chart, each representing $n$-th $s$ state, where the nuclei in those regions are likely to possess halo $s$-wave state near the neutron separation threshold, rendering large $\sigma_{\mathrm{TNC}}$. Due to periodic appearance of $ns$ states with different principal quantum numbers, those regions should appear repeatedly, just like `stripes' when represented in the nuclear chart as in \figref{NuChatMainFig}(a). Notably, the halo $s$-wave state and the Efimov states appear as excited states in the vicinity of the neutron separation thresholds. If we add an excitation energy as an additional axis on top of the nuclear chart as shown in  \figref{NuChatMainFig}(a), and represent the one- and two-neutron separation thresholds as the stability boundaries of the nuclei, then the $s$-wave halo state of $^{A+1}X$ and the Efimov state in $^{A+2}X$ appear in the vicinity of the curved surfaces representing $S_{1n}^{(\mathrm{GS})}$ and $S_{2n}^{(\mathrm{GS})}$, respectively. Consequently, the $s$-wave halo of $^{A+1}X$ (the Efimov states in $^{A+2}X$) are likely appear in the 3D stripe regions along the curved surfaces of $S_{1n}^{(\mathrm{GS})}$ ($S_{2n}^{(\mathrm{GS})}$), each corresponding to 2s, 3s, 4s... stripes. $^{90}$Zr and $^{159}$Gd belong to the 3s and 4s stripe, respectively. When we fix the number of neutrons and decrease the number of protons, moving toward the neutron-rich nuclei, the neutron separation energy decreases. It finally vanishes $S_{1n}^{(\mathrm{GS})}=0$ when approaching the neutron drip-line, so does the 3D stripes. The stripe regions along the curved surfaces of $S_{1n}^{(\mathrm{GS})}$ and $S_{2n}^{(\mathrm{GS})}$ therefore appear just like `ridges' along the mountains: starting their foots from the neutron-rich drip-line, increasing their heights up to plateaus of $S_{1n}^{(\mathrm{GS})}\sim 8$~MeV and $S_{2n}^{(\mathrm{GS})}\sim 16$~MeV for nuclei in the valley of stability. Notably, the ridge stripes for the $s$-wave halos of $^{A+1}X$ on $S_{1n}^{(\mathrm{GS})}$ as well as the Efimov states of $^{A+2}X$ on $S_{2n}^{(\mathrm{GS})}$ are likely to appear in a parallel direction along the proton number axis. This is because the change of $S_{1n}^{(\mathrm{GS})}-E_x$ caused by the change of the neutron number tends to be much more significant than the change of proton number; if the neutron number is changed, the Fermi energy of the neutron gets dramatically changed, leading to significant change of $E_x$. By contrast, if the proton number is changed with a fixed neutron number, then the energy levels of the neutrons shift. Unless this shift is so significant that the nuclei suddenly change the neutron occupation, the changes in $S_{1n}^{(\mathrm{GS})}$ and $E_x$ are compensated by each other, resulting in a minor change in $S_{1n}^{(\mathrm{GS})}-E_x$. We therefore arrive at the schematic picture in \figref{NuChatMainFig}(a), where the Efimov states likely appear in the ridge stripes regions, straddling above the nuclear chart and valley of stability, along the $S_{2n}^{(\mathrm{GS})}$ surface.


The ridge stripes picture can naturally encompass conventional candidates of the $s$-wave halo and the Efimov states in the neutron-rich nuclei: the halo nuclei close to the neutron drip-line has a small $S_{1n}^{(\mathrm{GS})}$. In most candidates for Efimov states, the $ns$ state should be close to the threshold. Therefore, these nuclei can be considered as belonging to the ridge stripe for the $ns$ state around the foot of the ridge $S_{1n}^{(\mathrm{GS})},S_{2n}^{(\mathrm{GS})}\approx 0$. As specific examples, $^{11}$Li and $^{19}$B, $^{20}$C, and $^{22}$C~\cite{FREDERICO2012939,canham2008universal,ACHARYA2013196}, may be qualitatively considered as belonging to the edge of the 2$s$ ridge. A super-neutron-rich nuclei $^{62}$Ca and $^{72}$Ca is theoretically studied as a candidate for the Efimov-like halo states~\cite{PhysRevLett.111.132501,PhysRevC.65.041302,PhysRevLett.120.052502,PhysRevC.105.024310}, which can be considered as belonging to the 3$s$ ridge.











\section{\label{sec:concl}Conclusions \& Outlook} 
We studied nuclei excited in the vicinity of the neutron separation threshold, and proposed a systematic way to search for nuclei with a large $s$-wave scattering length using the thermal neutron capture cross-section data. The candidate nuclei found by our protocol, $^{88}$Zr and $^{157}$Gd, have enormous neutron capture cross-sections, suggesting doorway halo states with gigantic spatial sizes and hence extremely large $s$-wave scattering lengths, which are 1-2 orders of magnitude larger than those in previous low-energy nuclear studies. Because of the large $s$-wave scattering length, these nuclei are ideal testbeds for studying Efimov physics in nuclear systems. We demonstrate that at least one, and possibly two, three-body bound states of a nucleus plus two neutrons of Efimov nature appear as the sub-threshold excited states in $^{90}$Zr and $^{159}$Gd. These are not the only candidates, but rather a few examples of richer varieties of halo nuclear states that appear in the vicinity of the neutron separation threshold. We have argued that the $s$-wave sub-threshold halo states and the Efimov states should appear in the `ridge stripes' regions excited above the nuclear chart as schematically depicted in \figref{NuChatMainFig}, each stripe region representing $ns$ valence state ($n=2,3,4,...$). Our ridge stripes picture naturally encompasses both the conventional Efimov candidates of neutron-rich nuclei as a special case of $S_{1n},S_{2n} \rightarrow 0$, and our newly proposed Zr and Gd in the middle of the valley of stability as $S_{2n} \approx 16$~MeV.

Further investigations into the properties of the sub-threshold excited states are necessary to unveil the nature of the nuclear Efimov states: e.g. their structures and reactions, detailed many-body calculations with more sophisticated interactions. $^{90}$Zr and $^{159}$Gd are good candidates to start exploring the nuclear Efimov physics in the experiments, and will open a novel avenue for studies on the halo nuclei and their universality. In particular, we expect their signatures can be observed from an inelastic scattering or a two-neutron transfer reaction; because $^{90}$Zr is a stable nucleus, an inelastic scattering may excite the ground state to the Efimov states. Alternatively, the Efimov states in $^{159}$Gd can be observed by a two-neutron transfer reaction on a stable nucleus $^{157}$Gd. The other nuclei with a large neutron capture cross-section such as $^{135}$Xe~\cite{PhysRev.102.823,shusterman2019surprisingly} would also be a notable candidate as shown in \figref{NuChatMainFig}. However, it is more challenging owing to unstable nature of $^{135}$Xe and  $^{137}$Xe, similarly as unstable $^{88}$Zr, $^{159}$Gd nuclei. We anticipate that the Efimov states in such nuclei may be observed in the near future nuclear experiments with radio-isotope beams.




The halo nuclear states are indispensable building blocks to understand the origin of the elements and the stellar reaction in the universe, and are at the frontier of the nuclear studies. In addition to the neutron-rich nuclei, proton-rich nuclei close to the proton drip-line are equally relevant. There are even more exotic cluster nuclei with $\alpha$ particles:$^{8}$Be, which is considered as a two-body bound state of two $\alpha$ particles with a large $s$-wave scattering length in the vicinity of $\alpha$ breakup threshold. The Hoyle state~\cite{hoyle1954nuclear}, an excited state of $^{12}$C, has been speculated as a three-body Efimov-like state of three $\alpha$ particles. In contrast to the sub-threshold states around the neutron separation threshold, these states are significantly affected by the Coulomb interaction, some of them becoming resonant states pushed up into the continuum. Although the relationship between these charged halo nuclei and our excited halos need to be clarified, our study has added a new avenue, with specific candidate nuclei, for the halo nuclear physics investigations. Furthermore, there are possibilities of $p$-wave or $d$-wave sub-threshold excited states, which may also be explored by looking at the thermal neutron capture cross-section at small but moderately large incident energy. The $p$-wave and $d$-wave halos may also appear as the ridge stripes of $np$ and $nd$ states above the nuclear chart, interweaving with each other and with the $s$-wave stripes presented in this paper.


\clearpage









\clearpage

\begin{acknowledgments}
 We thank Tokuro Fukui, Tomohiro Uesaka, Susumu Shimoura, Kazuyuki Ogata, and Emiko Hiyama for fruitful discussions. This work was supported by JSPS KAKENHI Grant Numbers JP21H00116, JP22K03492, JP23H01174,  and also by the RCNP Collaboration Research Network program as the project number COREnet-050.
\end{acknowledgments}

\appendix
\section{\label{sec:method}Methods}

\subsection{Three-body model}

We consider a three-body problem of a core nucleus plus two neutrons in the opposite spin states.  We assume that the interaction potential between the core nucleus and neutrons are the same for spin-up and spin-down neutrons. This is particularly true for $^{88}$Zr ($0^+$ state) owing to its symmetry, but a working hypothesis for $^{157}$Gd  ($3/2^-$ state) owing to the lack of nuclear data. The interaction between the nucleus core and a neutron is modelled as the Woods--Saxon potential
\begin{equation}
  V(r)= \frac{V_0}{1+\exp\left(\frac{r-R}{r_0}\right)},
  \label{eq:WS_pot}
\end{equation}
with $R= 1.25A^{1/3}$~fm, $r_0 = 0.65$~fm, and $A=88, 157$ is the mass number for Zr and Gd, respectively. The depth of the potential $V_0$ is varied to change the $s$-wave scattering length $a$ between the core nucleus and the neutron for obtaining the Efimov spectrum as a function of $a^{-1}$. More specifically, the depth of $V_0$  is tuned to be close to third and fourth $s$-wave resonances, which correspond to the dominant threshold 3s and 4s orbitals in $^{90}$Zr and $^{159}$Gd, respectively. Indeed, the selected $V_0$ are in the range of $- 35\textendash65$~MeV and $- 45\textendash75$~MeV respectively, which are consistent with the realistic Woods--Saxon potential depth $V_0= - 40\textendash50$~MeV~\cite{RingShuck}. For the interaction between the neutrons in the opposite spin states, we adopt the AV4 potential with $a_{nn} = -19.5$~fm~\cite{PhysRevLett.89.182501} as a realistic neutron-neutron potential in \figref{AllSpeciesPhysicalRes} because we can essentially focus on the central part of the interaction at low energy. To compare it with a idealized neutron-neutron interaction strictly at the unitary limit $a_{nn} = \pm \infty$,  we also perform the calculation with a scaled  AV4 potential $\lambda V_{nn}(r)$: $\lambda$ is fine-tuned to be $a_{nn} = \pm \infty$ around the first $s$-wave resonance, which renders $\lambda = 1.07....$ for the AV4 potential.

In solving the three-body problem, we use the separable potential method as introduced in Refs~\cite{pascaleno3BP1,pascaleno3BP2}. Namely, we construct separable functions $g |\chi \rangle \langle \chi |$ to exactly reproduce the zero-energy two-body wavefunctions for the Woods--Saxon and neutron-neutron potentials, respectively, and solve the Skorniakov--Ter-Martirosian-like equation. This method does not require an introduction of an artificial three-body parameter, but the scales of the Efimov spectrum is naturally set by the short-range correlations introduced in the separable functions.

\subsection{Different neutron-neutron interaction}

\begin{table}[!t]
\caption{Bound-state trimer's energy measured from the $^{88}$Zr +2n threshold $E_n^{(\mathrm{3B})}$, estimate of the size of the trimer $\sqrt{\langle r^2 \rangle }\sim \sqrt{\hbar^2/2m_n|E_n^{(\mathrm{3B})}|} $, and the ratio between the adjacent trimers are shown for $^{90}$Zr for three different types of neutron-neutron potentials: AV4~\cite{PhysRevLett.89.182501}, Minnesota~\cite{THOMPSON197753}, and CDMT13~\cite{PhysRevC.100.011603}. We also show the results for a hypothetical neutron-neutron interaction $\lambda V_{nn}(r)$, where $\lambda$ is fine-tuned to be at the unitary neutron-neutron interaction $a_{nn=}\pm \infty$. The $s$-wave scattering length between $^{88}$Zr and a neutron is estimated from the thermal neutron capture cross-section (\equref{TNCaEstimate}) as  $|a|=5.1(2) \times 10^3$~fm. The error originates from numerical uncertainty $\lesssim 0.5\%$ and the ambiguity in the sign of $a$. The universal ratio dictated by the universal Efimov theory is $e^{\pi/|s|}=15.95...$.}
\centering
\label{tab:result_list_Zr} 
\begin{tabular}{ccccc}
\hline
Potential&$a_{nn}$[fm]&$|E_n^{(\mathrm{3B})}|$[keV]&$\sqrt{\dfrac{\hbar^2}{2m_n|E_n^{(\mathrm{3B})}|}}$[fm]&$\sqrt{\dfrac{E_n^{(\mathrm{3B})}}{E_{n+1}^{(\mathrm{3B})}}}$\\
\hline
\multirow{6}{*}{AV4} & \multirow{4}{*}{$\pm \infty$} &$466(2)$& 6.67(3)& 2.39(1)\\
 & &$81.7(4)$&15.9(1)& 15.3(5)\\
 & &$0.351(21)$& 243(7)&19(9)\\
 & &$1.8(1.3)\times 10^{-3}$ & $5(3)\times 10^3$ & -- \\ \cline{2-5}
 & \multirow{2}{*}{$-19.5$} & $176(1)$ & $10.9(1)$ & $14.3(1)$\\
 & & $0.87(1)$ & $154(1)$ & --\\
\hline
\multirow{6}{*}{Minnesota} & \multirow{4}{*}{$\pm \infty$} & $292(1)$ & 8.42(4) & 1.97(1)\\
 & & $75.3(3)$ & 16.6(1) & 15.6(5)\\
 & & $0.311(20)$ & 259(9) & 20(9) \\
 & & $1.6(1.2)\times 10^{-3}$ & $5.2(2.5)\times 10^3$& -- \\ \cline{2-5}
 & \multirow{2}{*}{$-16.9$} & $70.6(4)$ & $17.1(1)$ & -- \\
 & & $\le 1.3 \times 10^{-3}$ & $\ge 4.1 \times 10^3$  & --\\
 \hline
\multirow{6}{*}{CDMT13} & \multirow{4}{*}{$\pm \infty$} & $305(2)$ & $8.23(4)$ & 2.06(1)\\
 &   & $72.2(3)$& $16.9(1)$ & 15.5(5) \\
 &   & $0.302(20) $ & $262(9)$ & 20(9) \\
 & & $1.5(1.2)\times 10^{-3}$ & $5.4(2.7)\times 10^{3}$ & -- \\ \cline{2-5}
 & \multirow{2}{*}{$-18.59$} & $81.2(4)$ & $16.0(1)$  & --\\
 & & $\le 1.4\times 10^{-3} $ & $\ge 3.8\times 10^3$  & --\\
 \hline
\end{tabular}
\end{table}


To understand how universal our results are with respect to inter-particle interactions, we also perform our three-body calculations with two other neutron-neutron interactions: the Minnesota two-range Gaussian potential with $a_{nn} = -16.9$~fm~\cite{THOMPSON197753}, and charge-dependent Malfliet Tjon two-range Yukawa potential (CDMT13)~\cite{malfliet1969solution} which are adapted to well-reproduce the low-energy scattering characteristics of the neutrons $a_{nn} = -18.59$~fm~\cite{PhysRevC.100.011603}. These potentials have different values $s$-wave scattering lengths, in addition to behaving rather differently at short distance. The results are summarized in Tables~\ref{tab:result_list_Zr} and~\ref{tab:result_list_Gd} together with those of AV4 potential discussed in the main text and in \figref{AllSpeciesPhysicalRes}. First, the precise values of the binding energies are different for different potentials owing to the difference in the $s$-wave scattering lengths and short-range form of the potentials. The latter effect can be interpreted as changing the value of the three-body parameter, which affects the absolute value of the binding energy of the Efimov states while the other universal features remain unaffected. The first and second excited states $E_2^{(\mathrm{3B})},E_3^{(\mathrm{3B})}$ of the unitary limit case $a_{nn}=\pm \infty$ vary by $\sim 15\%$, which is comparable to the variance of the three-body parameter found in the case of identical bosons~\cite{pascaleno3BP1,pascaleno3BP2}. In the other respects, the trimer energies for different nn interactions with $a_{nn}=\pm \infty$ behaves almost perfectly in the same manner, with higher excited states well reproducing the scaling ratio as dictated by the universal Efimov theory. The second and third excited states in Tables~\ref{tab:result_list_Zr} and~\ref{tab:result_list_Gd} seem to have large errors due to the uncertainty in the sign and value of the $s$-wave scattering length estimated from the thermal neutron capture cross-section data. If we compare $E_n^{(\mathrm{3B})}$ strictly at $1/a=0$, we find no such errors and the higher excited states better reproduce the universal scale factor.

The difference in the value of $s$-wave scattering length $a_{nn}$ has much more significant impact because the number of the Efimov states gets smaller for smaller value of the $s$-wave scattering length $|a_{nn}|$. For the AV4 potential, there are two trimer states for both $a_{nn}>0$ and $a_{nn}<0$ sides. The ratio of the binding energies of $^{90}$Zr is 14.3(1), which is in good agreement with the Efimov universal value 15.95, even though $a_{nn}$ and $a$ are not strictly at the unitary limit. This indicates the possibility of testing the discrete scale invariance of the Efimov states in nuclear physics; however, some remarks are in order. First, the binding energy of the excited state is $\lesssim$~keV, which is extremely small and a minute change in the value of the $s$-wave scattering length can significantly affect the presence of this excited Efimov state. For the Minnesota and CDMT13 potentials, we only find one trimer state. This is consistent with the number of the Efimov period estimated from the universal Efimov theory $\frac{|s|}{\pi} \ln(|a_{nn}|/R)\approx 0.4-0.5$: due to finite value of $a_{nn}$, there can be 1 or at best 2 Efimov states. The precise values of the binding energies of the trimer $E_1^{(\mathrm{3B})}$ is also affected by the values of $a_{nn}$, because its spatial size $\sqrt{\langle r^2 \rangle}$ is comparable with $|a_{nn}|$. While the precise values of the binding energies are affected by the detailed shape of the potential and $a_{nn}$, we find the other features of the tighter bound trimer state reamains the same as those of AV4 potential.

\begin{table}[!t]
\caption{Same as Tables~\ref{tab:result_list_Zr} for $^{159}$Gd. The $s$-wave scattering length between $^{157}$Gd and a neutron is estimated from the thermal neutron capture cross-section (\equref{TNCaEstimate}) as $|a|=2.84(1) \times 10^3$ fm. The universal ratio dictated by the universal Efimov theory is $e^{\pi/|s|}=15.86...$.}
\centering
\label{tab:result_list_Gd} 
\begin{tabular}{ccccc}
\hline
Potential&$a_{nn}$[fm]&$|E_n^{(\mathrm{3B})}|$[keV]&$\sqrt{\dfrac{\hbar^2}{2m_n|E_n^{(\mathrm{3B})}|}}$[fm]&$\sqrt{\dfrac{E_n^{(\mathrm{3B})}}{E_{n+1}^{(\mathrm{3B})}}}$\\
\hline
\multirow{6}{*}{AV4} & \multirow{4}{*}{$\pm \infty$} &556(3) & 6.11(3) & 4.17(4)\\
 & & 31.9(5) & 25.5(3) & 16(2)\\
 & & 0.125(23) & 412(38) & --\\
 & & $\le 3.3\times 10^{-3}$ & $\ge 2.5\times 10^3$ & -- \\ \cline{2-5}
 & \multirow{2}{*}{$-19.5$} & 261(2)&8.90(4)& 7.6(2)\\
 & & 4.51(24) & 67.8(1.8) & --\\
\hline
\multirow{6}{*}{Minnesota} & \multirow{4}{*}{$\pm \infty$} & 233(2) & 9.43(5) & 3.06(3)\\
 & & 24.8(2) & 28.9(2) & 15.6(1.4)\\
 & & 0.104(20) & 453(44) & -- \\
 & & $\le 3.1\times 10^{-3}$  & $\ge 2.6 \times 10^3$ & -- \\ \cline{2-5}
 & \multirow{2}{*}{$-16.9$} & 47.0(2) & 21.0(1) & -- \\
 & & $\le 3.7\times 10^{-3}$ & $\ge 2.4\times 10^{3} $  & --\\
 \hline
\multirow{6}{*}{CDMT13} & \multirow{4}{*}{$\pm \infty$} & 279(2) & 8.62(4) & 3.24(3)\\
 &   & 26.6(3)& 27.9(2) & 15.8(1.4)\\
 &   & 109(21) &  442(42) & -- \\
 & & $\le 3.1\times 10^{-3}$ & $\ge 2.6\times 10^3$ & -- \\ \cline{2-5}
 & \multirow{2}{*}{$-18.59$} & 65.3(3) & 17.8 & -- \\
 & & $\le 5.1 \times 10^{-3}$ & $\ge 2.0\times 10^3$ & --\\
 \hline
\end{tabular}
\end{table}

%


\end{document}